\documentclass[10pt,conference]{IEEEtran}
\IEEEoverridecommandlockouts

\usepackage{cite}
\usepackage{amsmath,amssymb,amsfonts}
\usepackage{algorithmic}
\usepackage{graphicx}
\usepackage{textcomp}
\usepackage{xcolor}
\def\BibTeX{{\rm B\kern-.05em{\sc i\kern-.025em b}\kern-.08em
    T\kern-.1667em\lower.7ex\hbox{E}\kern-.125emX}}

\usepackage{pbalance}

\makeatletter
\newcommand{\linebreakand}{%
  \end{@IEEEauthorhalign}
  \hfill\mbox{}\par
  \mbox{}\hfill\begin{@IEEEauthorhalign}
}
\makeatother

\begin{document}
\title{The Junction of Immersive Analytics and Virtual Reconstructions --
        A Case Study on the Mausoleum of Emperor Maxentius
\thanks{DFG, project ID 251654672 – TRR 161}
}

\author{%
\IEEEauthorblockN{Wilhelm Kerle-Malcharek}
\IEEEauthorblockA{\textit{Computer and Information Science} \\
\textit{University of Konstanz}\\
Konstanz, Germany \\
wilhelm.kerle@uni-konstanz.de}
\and
\IEEEauthorblockN{Niklas Hann-von-Weyhern}
\IEEEauthorblockA{\textit{Department of History} \\
\textit{University of Konstanz}\\
Konstanz, Germany \\
niklas.hann-von-weyhern@uni-konstanz.de}
\and
\IEEEauthorblockN{Ulf Hailer}
\IEEEauthorblockA{\textit{Department of History} \\
\textit{University of Konstanz}\\
Konstanz, Germany \\
ulf.hailer@uni-konstanz.de}
\linebreakand 
\IEEEauthorblockN{Steffen Diefenbach}
\IEEEauthorblockA{\textit{Department of History} \\
\textit{University of Konstanz}\\
Konstanz, Germany \\
steffen.diefenbach@uni-konstanz.de}
\and
\IEEEauthorblockN{Stefan P. Feyer}
\IEEEauthorblockA{\textit{Computer and Information Science} \\
\textit{University of Konstanz}\\
Konstanz, Germany \\
stefan.feyer@uni-konstanz.de}
\and
\IEEEauthorblockN{Karsten Klein}
\IEEEauthorblockA{\textit{Computer and Information Science} \\
\textit{University of Konstanz}\\
Konstanz, Germany \\
karsten.klein@uni-konstanz.de}
\linebreakand
\IEEEauthorblockN{Falk Schreiber}
\IEEEauthorblockA{
\textit{Computer and Information Science, University of Konstanz} \\
\textit{Faculty of Information Technology, Monash University} \\
Konstanz, Germany \\
falk.schreiber@uni-konstanz.de}
}
\maketitle

\begin{abstract}
Virtual archaeology has significantly evolved over the last few decades through advancements in data acquisition and representation; for example, by improved data recording technologies and virtual reality devices. Immersive environments provide novel ways to present historical events or objects with high visual quality for both the general public and researchers. Here, we examine how the emerging field of immersive analytics can contribute to enhancing the understanding and exploration of archaeological data, and we explore the junction of virtual archaeology and immersive analytics. We discuss a selection of features already used by the community and examine how optimizing these can facilitate the discourse on cultural heritage objects. As a basis for discussion, we introduce and utilize three digital reconstruction interpretations of the mausoleum of the late Roman Emperor Maxentius in Rome, which are based on prior scientific work and a typological framework. Based on our work, we advocate for the value of combining historical and computer science expertise to optimize immersive environments for virtual reconstructions, thereby facilitating a deeper understanding and interactive exploration of archaeological data. 
\end{abstract}

\begin{IEEEkeywords}
virtual archaeology; cultural heritage; 3D reconstruction; immersive analytics; virtual reality; Maxentius mausoleum;
\end{IEEEkeywords}

\section{Introduction}
Terms such as virtual archaeology, virtual restoration, and virtual reconstruction date back to the 1990s, when researchers investigated the potential of virtualization for work on cultural heritage~\cite{Reilly90,prescott1997electronic}.
Over the years, virtual archaeology has evolved into an entire field of research, with significant advancements in its technological capabilities in terms of data acquisition and data representation, as well as the development of charters that employ meaningful principles for visualizations and virtual reconstructions~\cite{lopez2011principles, denard2012new}.
According to the Seville Charter~\cite{lopez2011principles}, virtual archaeology concerns itself with finding comprehensive ways of visualizing cultural heritage objects.

The visualization of archaeological objects began in the 2D realm and has since increasingly seen an extension through 3D to stereoscopic 3D (s3D) applications (e.\,g., models in virtual reality) since 2D visualizations often fall short of capturing the complexity and spatial relationships inherent in archaeological data.
Immersive technologies can typically be classified by the reality-virtuality continuum by Milgram and Kishino~\cite{milgram1994taxonomy}.
Most commonly, immersive technologies are understood to be either virtual reality (VR), which is an entirely virtual environment; augmented reality (AR), which is an overlay of additional information into the real world; or mixed reality (MR), which is a blend between the virtual and real worlds.
By leveraging immersive technologies, researchers can generate detailed and realistic 3D experiences of not only the sites and artifacts but also reconstructed objects, background information, or a contextually appropriate environment.

Rome, as a focal point of archaeological work in Europe, enjoys high popularity for such work, leading to an increase in more extensive reconstruction visualizations.
An example of such an immersive model representation is \textit{Rome Reborn}, a virtual reality project that has been running since 1996 at the University of Indiana, which reconstructs Rome around 320 AD~\cite[pp.~23--25]{frischer20223d}. 
The transparency of the underlying archaeological data origin of such public audience-oriented applications tends to be selective, and their authenticity and reliability are therefore limited.
This is only natural because the goal of an immersive environment targeted at the broader public is not to discuss different scholarly opinions regarding the reconstruction~\cite[pp.~13--14]{liverani20213d}. 

To support scientific debates, digital reconstructions are useful for developing innovative questions about the historical perception and sociopolitical context of monuments and objects, visualizing different chronological phases of archaeological sites, and reflecting on the construction history and building process of historical monuments.
A clear indication of different phases, information sources, or certainty of validity can, among others, take the form of color mappings or opacity of the respective components integrated into the visualization~\cite{bartz2016digitales, demetrescu2017virtual, pietroni2021virtual}.

To combine the potential of immersive environments with scientific debates, the relatively young field of immersive analytics (IA) offers numerous advancements.
IA utilizes engaging and embodied tools to support data understanding~\cite{marriott2018immersive}, explores the potential of immersive environments for data analysis, and develops and evaluates designs of such environments tailored for specific application areas.
IA combines topics from visual analytics, computer graphics, human-computer interaction, and virtual reality.
As a consequence, several areas that exhibit potential for archaeological research, including s3D, collaborative analysis, and multimodal representation and interaction, can be considered a part of it~\cite{bienroth2022spatially,nowke2013visnest}.
The investigation of which navigation or interaction methods are most appropriate for a specific context~\cite{drogemuller2018evaluating}, how effective VR is in teaching contexts~\cite{sari2024cognitive}, or the impact of different modalities such as audio or haptics~\cite{ruotolo2013immersive, venkatesan2023haptic} are only some of its research topics.

In addition to enhancing data representation and analysis, IA can help communicate results in a more engaging and accessible way, such as by investigating how to annotate specific objects or positions or how to collaborate in which contexts. 
Creating immersive environments that allow people to virtually explore sites, artifacts, and reconstructed objects, add annotations and landmarks for idea communication, and obtain more than mere visual impressions of cultural heritage objects supports bridging the gap between academia and the public.
Thus, it is crucial to emphasize the optimization of those supporting features based on the cultural heritage item displayed and the user group towards which the virtual environment is targeted.

\begin{figure}[tb]
\centerline{\includegraphics[width=\linewidth]{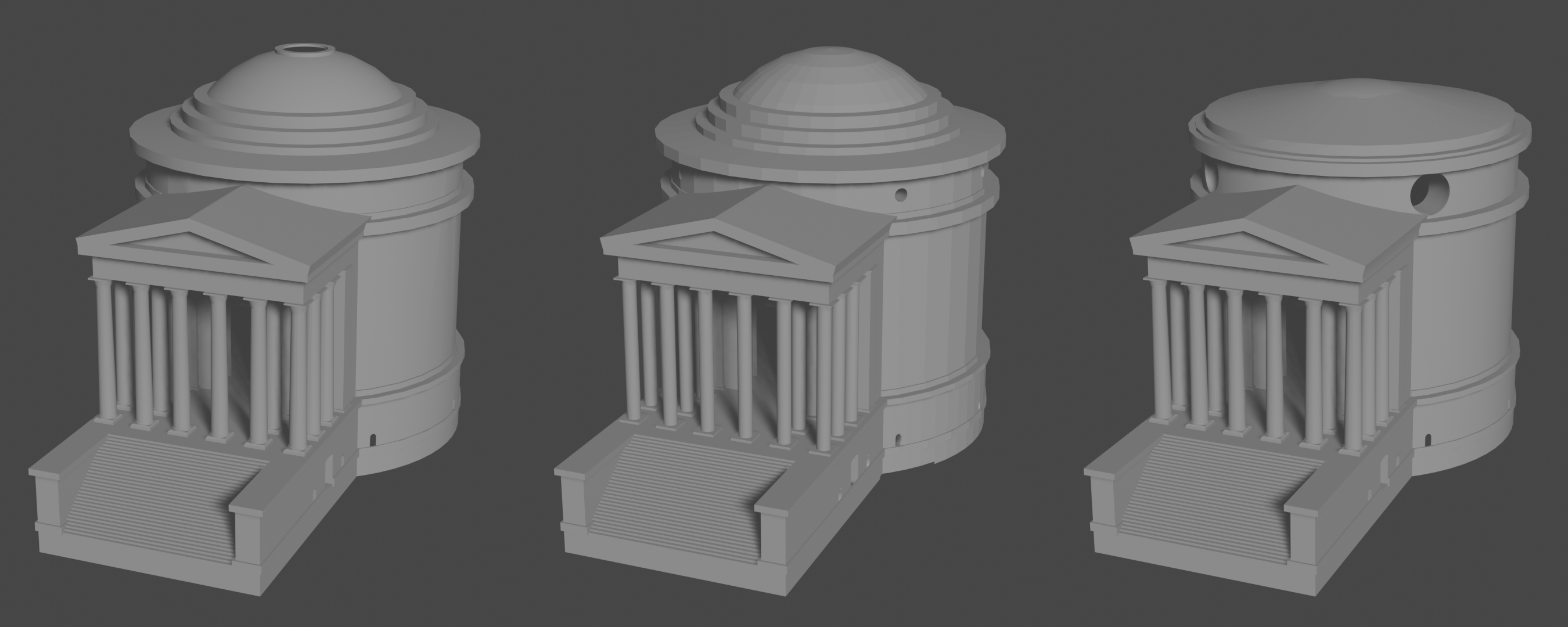}}
\caption{This image illustrates the three proposed models of the Mausoleum of Maxentius discussed in this work. The first model (Maxentius A) closely follows Rasch's reconstruction proposal~\cite{rasch1984maxentius}. The second model (Maxentius B) is based on Rasch's model but features a closed dome and seven windows in the clerestory, modeled after the Tor de' Schiavi~\cite{Johnson2009}. The third model (Maxentius C) attempts to merge the dome and windows of Tor de' Schiavi with the proportions of the Maxentius Mausoleum.}
\label{fig:ex3}
\end{figure}

Since immersive experiences are widely utilized for dissemination, such as in museums or teaching, as well as for research~\cite{boboc2022augmented,rodriguez2024systematic}, and evidence suggests that discourse in virtual archaeology is primarily driven by advancing technologies~\cite{munster2019digital}, we discuss perspectives on how the mentioned potentials can be utilized.
We consider how immersive analytics methods can be used for virtual archaeology, specifically for virtual reconstructions. 
We use three reconstructions of the Maxentius Mausoleum in Rome for our argumentation (see Fig.~\ref{fig:ex3}).
They are based on ongoing scientific debates about what the mausoleum might have looked like in its prime. 
We provide these reconstruction models, along with software, to inspect them in VR.

Thus, our contribution consists of a perspective on how immersive analytics research can assist archaeological discourse, as well as the three digital reconstructions of the Maxentius mausoleum that we created. Furthermore, we provide an example software to demonstrate some of the possible concepts and techniques.

\section{Background}
The popularity of immersive technologies, such as AR and VR, for museums or expositions, teaching, and even research in archaeological contexts has grown~\cite{boboc2022augmented, rodriguez2024systematic,frischer20223d,schofield2018viking,tscheu2016augmented}.
As we focus on reconstructions in this work and its interdisciplinary nature, it's important to clarify the different types of reconstructions in the digital realm.
\emph{Digital reconstruction} involves creating objects using computational tools, while \emph{virtual reconstruction} is the use of virtual models to recover a building or manmade object visually~\cite{lopez2011principles}.
\emph{Immersive reconstruction} extends this further by embedding objects in environments along Milgram's reality-virtuality continuum~\cite{milgram1994taxonomy}, such as AR or VR, to enhance user experience through interaction and context.

As is the nature of a reconstruction, it is possible that not all pieces of a monument, for instance, are known.
Researchers must complement the known parts with modeled ones, whose form and structure derive from various sources, such as paintings, folklore, and derivations from existing pieces.
These different sources need to be indicated, as Demetrescu describes in his work~\cite{demetrescu2017virtual}.
The sources of the reconstruction components, as well as the surrounding area and information on the object's purpose or its material, are all data that can be transported via various means, especially so with the help of immersive technologies.

Immersive technologies can support reasoning~\cite{elsayed2015using} by embedding data into the physical world~\cite{willett2016embedded}, the recall of data~\cite{mark2003there}, or spatial judgment accuracy~\cite{ragan2012studying}.
This is also relevant in virtual archaeology, as such environments have proven to create a heightened sense of presence, as already shown by Kyrlitsias, Christofi, Michael-Grigoriou, Banakou, and Ioannou~\cite{kyrlitsias2020virtual}. 
Presence, as a form of \emph{psychological immersion}, can also be called \emph{engagement}~\cite{Bueschel2018}, and is an important concept, as there is evidence that task performance and recall are boosted through heightened engagement~\cite{makowski2017being}.

The research direction that concerns itself with investigating how to exploit tools which engage the user to increase data understanding is called \emph{immersive analytics}~\cite{marriott2018immersive}.
In their book on immersive analytics, Marriott et al. describe several additional relevant aspects apart from those already discussed.
One of these is \emph{situated analytics}, which places data representations in relation to relevant objects, places, and persons to enhance understanding and decision-making~\cite{ElSayed2015}.
For example, it might overlay an existing object with analysis information about it and is thus tightly linked to AR applications.
Another is the \emph{embodiment}, which relates to the interaction of a human with a system through touch, gestures, or voice as a form of physical affordance, rather than the point-and-click interaction typical of desktop PCs~\cite{Bueschel2018}.
That said, the concept of \emph{multi-modality} is closely related, as it encompasses input and output addressing multiple senses, such as vision, hearing, touch, kinesthesis, and (less commonly) smell.

\section{Maxentius Mausoleum}

\begin{figure}[tb]
\centerline{\includegraphics[width=\linewidth]{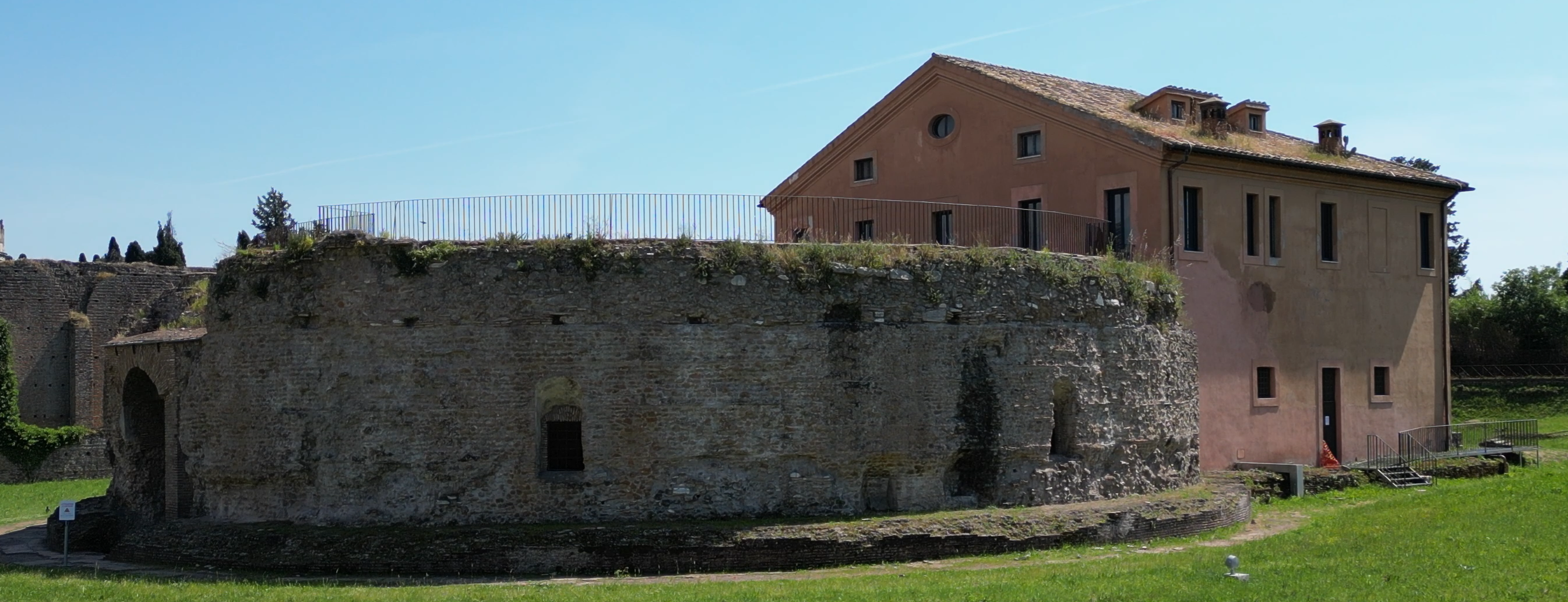}}
\caption{The current state of the Maxentius mausoleum. Only the crypt still remains. The front porch is partly missing and has partly been used to build a farmhouse.}
\label{fig:remains}
\end{figure}

To provide insight into the significance of our archaeological example, we briefly discuss its historical and typological embedding and connection to the reign of Maxentius.
The latter is relevant, as after nearly a century, during which the Roman emperors had been virtually absent from Rome and spent most of their time campaigning at the frontiers of the empire, the short rule of Maxentius (r.~306-312) gave Rome again the status of a permanent imperial residence. 
Maxentius styled himself as the conservator of Rome (conservator urbis suae) and added several large buildings such as the Maxentius Basilica and the temple of Venus and Rome, thermal baths, and an extension of the imperial palace on the Palatine Hill to the architectural ensemble of the city~\cite[pp.~12-13, 51, 74-82, 87-95]{leppin2007maxentius}.

On the city's outskirts, at the third milestone of the Via Appia leading southwards, Maxentius built a residential complex that encompassed a villa suburbana with an attached well-preserved circus and a monumental mausoleum facing the Via Appia. 
The mausoleum, which forms the core of the complex, is surrounded by a slightly offset portico of 99.50 by 85 meters, of which substantial remains (as high as 11 meters) can still be seen~\cite[p.~14]{rasch1984maxentius} (see Fig.~\ref{fig:remains}).
However, the preserved remains of the mausoleum give the modern visitor only a limited idea of its former importance and structure.
Just the foundations of the original building remain, overbuilt partly by a farmhouse of the 18$^{\text{th}}$ century~\cite[pp.~5--6]{rasch1984maxentius}.
Inside the building, an ambulatory crypt around an overarching pillar with niches inside is well-preserved. 
The building's layout features a circular podium with a diameter of 33.25 meters.
Adjacent is a front porch partly overbuilt by a farmhouse with a length of 17.00 meters and a width of 22.68 meters. 
Covered by the farmhouse, a vaulted cellar of the front porch with a diameter of 9.10 by 16.33 meters remains. 
The still-existing crypt has a diameter of 23.54 meters, a central column 9.70 meters wide, and seven niches~\cite[pp.~86--90]{Johnson2009}.

The monumental mausoleum was strongly connected to Maxentius' rule and likely almost completed when Maxentius was killed in 312 AD~\cite[pp.~92--93]{Johnson2009}.
Its few remaining traces provide an excellent case study for the application of virtual archaeology.
Typologically, the mausoleum can be attributed to a type of podium round building, which -- according to the architectural historian Jürgen Rasch -- ultimately derives from the Pantheon.
From 1977 to 1979, Rasch conducted an architectural survey of the site of the Maxentius Mausoleum and proposed a reconstruction of the building modeled on the Pantheon with its specific architectural components: a podium with a hexastyle front porch (pronaos), a domed circular building with seven niches, and a circular dome opening (oculus) for the second storey~\cite[p.~84]{rasch1993mausoleum}.
However, this Pantheon version of the Maxentius Mausoleum has been questioned by Mark Johnson, referring to a temporally and typologically similar building – another late antique imperial mausoleum at the via Praenestina (the so-called Mausoleum of Tor de' Schiavi)~\cite{rasch1993mausoleum}.
This monument is located, like the Maxentius Mausoleum, on the outskirts of Rome, close to an imperial villa suburbana. It follows the design of a podium round building featuring a slightly smaller, tetrastyle front porch with a domed vault without an oculus but four windows in the clerestory.
These different perspectives, into which Maxentius' Mausoleum can be classified, result in different conceivable models for its reconstruction, which will be discussed below.

\section{Immersive Analytics for the Maxentius Mausoleum}

\begin{figure}[tb]
\centerline{\includegraphics[width=\linewidth]{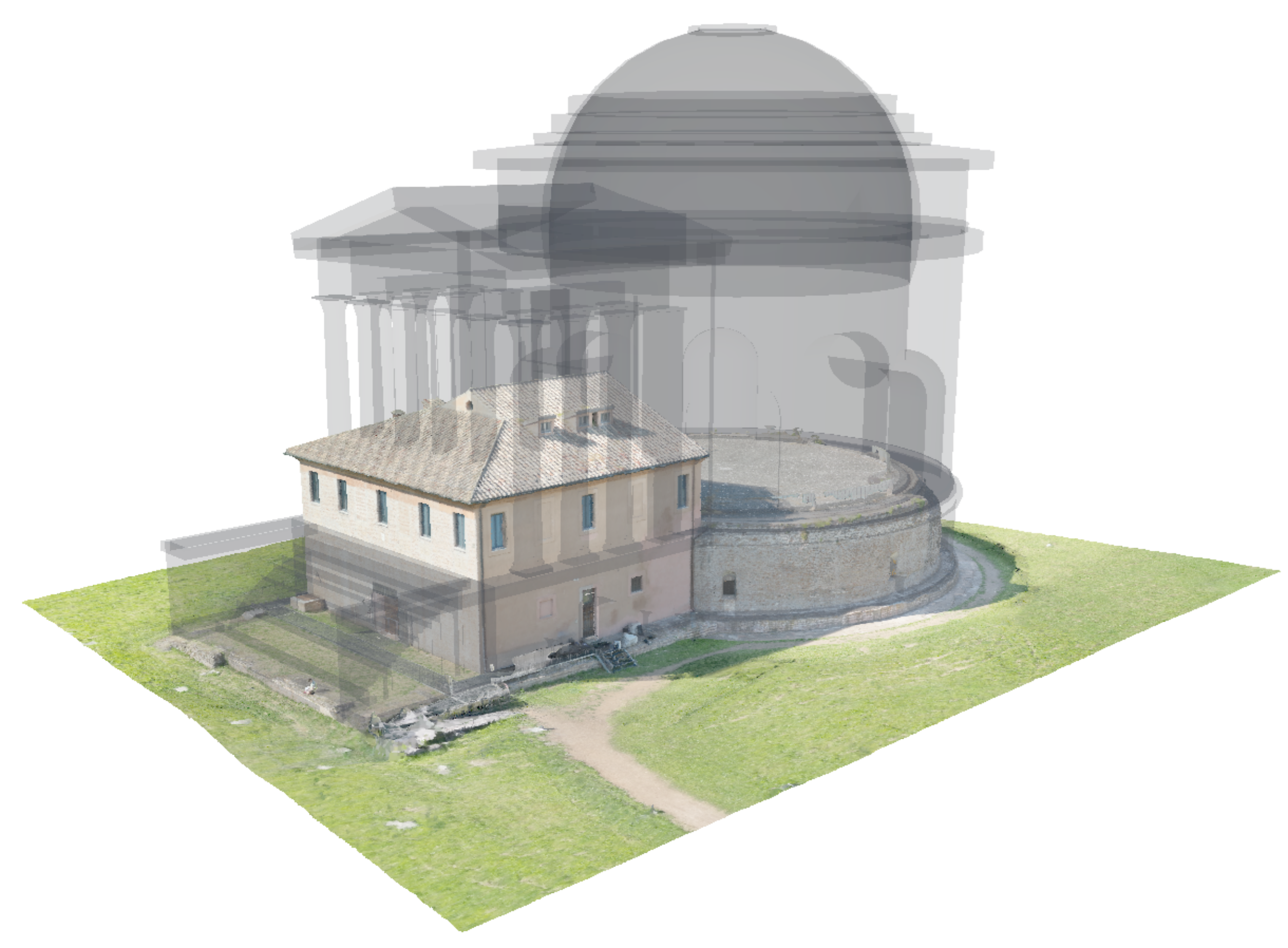}}
\caption{Cross-faded version of Maxentius A. The different parts of the building are visible, and a photogrammetric recording of the current state of the crypt is embedded into the reconstruction.}
\label{fig:maxA}
\end{figure}

Regarding the Maxentius Mausoleum, we chose Rasch's work as a foundation for a first reconstruction (Maxentius A), which closely follows his suggested measurements.
The second model (Maxentius B) assumes a closed dome following Johnson's proposal~\cite[pp.~91-92]{Johnson2009} and adds seven round windows in the clerestory with a diameter of 1.80 meters considering the proportions of Tor de' Schiavi~\cite[p. 28]{rasch1993mausoleum}.
The third model (Maxentius C) adapts the flatter dome of Tor de' Schiavi~\cite[p.~68]{rasch1993mausoleum} and the windows with a larger diameter.
This model visualizes, as suggested by Johnson, a version of the Maxentius Mausoleum modeled closely on Tor de' Schiavi~\cite[p.~92]{Johnson2009}.
Consequently, model B can be considered a merger of models A and C.
Thus, we have three reconstruction models of the same object (see Fig.~\ref{fig:ex3}), which mainly differ in the design of the dome, allowing us to discuss which of the hypotheses about that part of the mausoleum is most plausible.
We created the reconstructions with the 3D modeling software Blender.
The components of the models consist of combined simple geometric forms, which were manipulated through manual node shifting to obtain necessary details.
For each sub-component, once constructed from primitives, the measurements for those components were linearly scaled and applied.
Thanks to the precise measurements obtained from Rasch's work, the models provide correct proportions, which help us obtain a fitting impression of the mausoleum's former glory.
In Fig.~\ref{fig:maxA}, we demonstrate this via overlaying Maxentius A transparently over a photogrammetry-based model of the mausoleum acquired from airborne imaging using a drone. 

\begin{figure}[tb]
\centerline{\includegraphics[width=\linewidth]{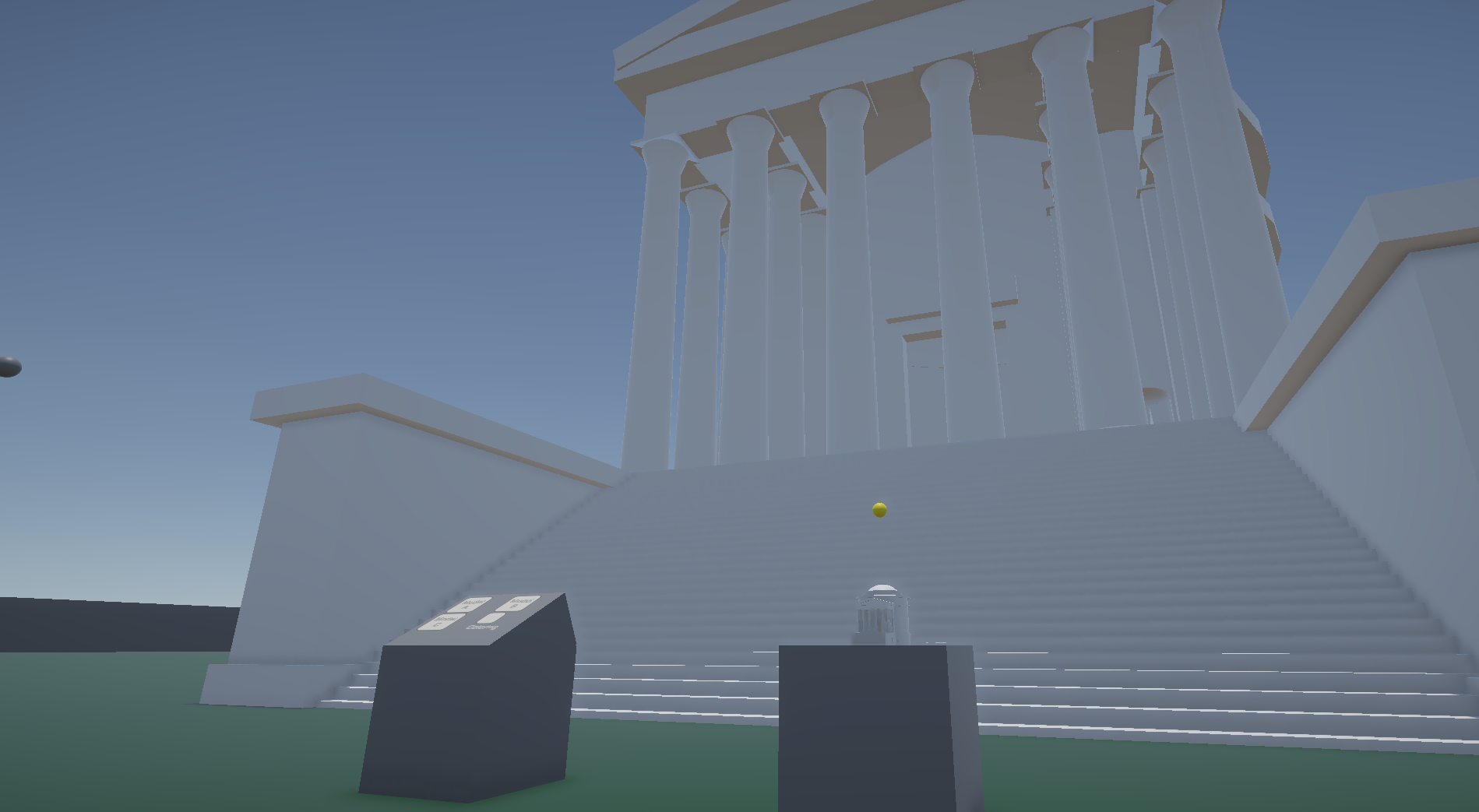}}
\caption{The software we provide together with one of our models. The panel, the miniature, and the actual visualization can be seen.}
\label{fig:softwareImage}
\end{figure}

We provide software (see Fig.~\ref{fig:softwareImage}) that presents the three hypothesized models of the Mausoleum of Maxentius.
The software provides a VR environment, implemented with Unity C\# and built to be used with an Oculus Quest 3 or Valve Index.
We chose VR as it represents an s3D environment with head-tracking, which allows for improved spatial judgment. Inside the immersive environment, the user will start in front of the Maxentius A stairway.
The user can move around, using the continuous movement metaphor, and can snap-turn to rotate. 

There is a panel with buttons to switch between the three reconstruction models and a toggle to turn on or off a coloring akin to the extended matrix by Demetrescu.
Users can pick up this panel and carry it through the VR scene to change the representation at any position.
Switching between models is especially relevant in a research context for comparing several different hypotheses directly while still in the iVR environment.
Consequently, in dialogue with domain experts, compiling ways to compare the reconstruction models intuitively or conveniently helps to improve cognitive affordance.

To the right of the panel is a miniature of the model.
By holding their virtual hands inside the miniature and performing a grabbing motion, a vibration of the controller indicates interaction, allowing them to control the angle of the light on the model.
There is a small yellow sphere indicating the relative position of the light source to the miniature.
This small sphere is intended to serve as a visualization to ease the intuition of where the light comes from.

This idiom is intended to aid in analyzing the source and distribution of sunlight for the mausoleum.
For our mausoleum scenario, it would be beneficial to explore ways to effectively interact with light sources. 
An optimized light interaction system could lead to a more fluent experience, which in turn may help experts justify the different shapes of the dome and the light conditions inside. 
Furthermore, users can place small squares to leave small randomly colored markers at points of interest behind.
In general, different types of markers could be developed.
For example, users could be enabled to colorize specific areas of the wall to mark the locations of the numerous drawings on the walls, which were added long after the mausoleum was originally built.
The type of marker, color coding, or even an acoustic signal could then be investigated for its effectiveness in locating the different spots. 

Lastly, the software can also be used without a VR headset, utilizing very simple keyboard and mouse controls, allowing it to be viewed without the need for additional hardware.
We provide the software via a Zenodo upload at~\cite{kerlemalcharek202514628075}.

\section{Discussion}
With this work, we present three digital reconstructions of the Maxentius Mausoleum in Rome and utilize them to assess the value of IA for virtual archaeology.
The reconstructions we propose for the Maxentius Mausoleum adhere to the typology of late antique imperial mausolea, based on archaeological evidence and scholarly interpretations. 
Such scientific evidence provides a sound foundation for our discussion, in which we aim to advocate for the more prominent use of immersive technologies to support archaeological research.
While there exist works that highlight the usefulness of immersive environments in virtual archaeology~\cite{kyrlitsias2020virtual}, they tend to focus on dissemination or reconstruction rather than improving the perception of a specific aspect, such as the comparison of hypotheses. 

From a historical and archaeological perspective, the immersive reconstruction of Maxentius’ mausoleum provides important insights into the changing relationship between light and space, a topic much debated in discussions on the development of Late Antique architecture.
As an outstanding example of a new type of two-storeyed mausolea combining a dark crypt with light-flooded upper floors, it adds a completely new dimension of including light to funeral architecture that was taken up and further developed in the fourth century by church-attached imperial mausolea~\cite{Hesberg1992}.
The graded lightning seen here became an important feature of late antique interiors, extending from funeral to palatial~\cite{gunter1968wand} and ecclesiastical~\cite{deBlaauw2010} architecture.
An immersive reconstruction of Maxentius’ mausoleum complex will explore this dimension of light and lightning in a systematical manner and, at the same time, offer a promising starting point to take the general discussion of this important feature of late antique architecture to a new level. 

Thus, using the three reconstruction models, we created immersive reconstructions in the form of an interactive VR (iVR) environment.
This iVR environment serves as proof of concept of the potential of IA in archaeological analysis. 
We integrated a selection of IA techniques to demonstrate a promising approach to accessing archaeological data that goes beyond representation.
Our environment enables users to manipulate lighting conditions, toggle between color-coded representations of reconstruction certainty and grey, place spatial markers, and navigate with a continuous movement metaphor.
They use their hands to adjust the lighting and gain haptic feedback when using buttons within the virtual space.
We chose those interactions to align with IA research suggesting that embodied tools, among other concepts, can improve data comprehension~\cite{marriott2018immersive}.
We chose VR because it enables interaction with the reconstruction in an immersive environment, supporting a deeper understanding of spatial relationships.
This finding is consistent with findings on the benefits of stereoscopic 3D and head-tracking for spatial judgment accuracy~\cite{ragan2012studying}.
The color coding we use builds upon established practices in virtual archaeology to represent reliability in reconstructions~\cite{demetrescu2017virtual}.

As the feature space of IA topics and virtual archaeology is vast, we decided against integrating an exhaustive number of potential features into our software.
Our objective was not to provide comprehensive coverage but to demonstrate a number of possibilities as an example. 
The interaction possibilities we offer allow observers to gain a general impression of the reconstruction alternatives by switching between them on demand.
Furthermore, observers can gain an intuition of the relative proportions of known parts of the mausoleum versus those that are not known but can be derived from similar architecture.
Through the miniature, users can see how the position of the light source changes relative to the reconstruction.

Due to the nature of this work, several limitations should be considered.
As for the models we created, as is usual in the domain, they also have parts that are more interpretative than others. 
The current implementation is focused on a single site and a specific set of interaction techniques.
Although multi-modality is important, only haptic feedback and a visual representation of the models are available.
Also, the use of VR hardware limits accessibility for some users.
While we do offer simple controls that allow the software to be used without VR hardware, the immersion factor that comes with VR is lost.
Additionally, the considerations we make involve several potential intersections of the two fields, without taking into account the in-depth considerations of how the design space of different methods, such as interaction design, would need to be applied.

For future work, a unified framework for creating immersive experiences would pose an asset.
Such a framework should encompass the inclusion and on-demand availability of historical context within an immersive environment.
This could happen in the form of integrated analysis tools, such as classifiers for identifying anachronistic components within a digitization. 
The sonification of information may also be of interest. 
It may also be helpful to have ways of adjusting the light conditions tailored to the object's geolocation and a chosen time, date, and seasonal characteristics, which was out of the scope of this work.

However, not only software is required.
A relevant direction is to assess specific and established workflows from IA and how they can be transferred and used in virtual archaeology.
It could also be valuable to compare the effectiveness of immersive environments with traditional methods of archaeological data analysis by evaluating the performance of experts on related tasks, such as identifying archaeological units.
Lastly, it would be interesting to see the effect of new metaphors for representing inherent information, such as levels of certainty, volumes of objects, sizes, usage, or chronological embedding.

Therefore, there are numerous opportunities in the future at this intersection, which should be explored and evaluated together with domain experts to gain further insights into the use of the Maxentius Mausoleum, specifically, and cultural heritage objects in general.

\section{Conclusion}
In this work, we present three reconstruction hypotheses of the Maxentius Mausoleum, along with the literature and typology that inspired them.
We created 3D models of those reconstruction hypotheses and provide them openly.
In addition, we have created an immersive environment in which these reconstructions can be viewed and explored through a small number of integrated interaction and navigation options, either as VR or desktop applications.
Furthermore, we explore the potential for virtual archaeology that immersive analytics as a research direction yields.
We advocate the shift from ever-better representations, recording, and storage of cultural heritage items to tailored immersive experiences that improve accessibility and insight into existing data.
The goal is to look at how virtual archaeology studies can be optimized by identifying the specific requirements of a given problem, such as comparing different monument reconstruction interpretations and investigating support options.
For argumentation, we outline the possibilities of engaging representations of reconstructions and interactive representations of hypotheses.
We exemplify the concept of integrating IA with our current research on the Maxentius Mausoleum, where these concepts will facilitate the analysis of hypotheses regarding its construction, use, and the relationship between light and space within it.
Overall, we present an exciting research direction that offers much potential for enhancing users' experience of virtual archaeology environments.

\section*{Acknowledgment}

We acknowledge funding by DFG, project ID 251654672--TRR 161 as well as the support of Antonella Bidlingmaier, who contributed to the creation of the first iterations of the reconstructions of the Maxentius Mausoleum.

\bibliographystyle{IEEEtran}
\bibliography{bib}

\end{document}